\documentclass[aps,prl,twocolumn,superscriptaddress,showpacs,floatfix]{revtex4-1}

\usepackage{graphicx}
\usepackage{dcolumn}
\usepackage{bm}
\usepackage{times}
\usepackage{color}
\usepackage{graphicx}
\begin{document}
	\bibliographystyle{apsrev}
	\title{Interlayer quantum transport in Dirac semimetal BaGa$_2$}
	
	\author{Sheng Xu}\thanks{These authors contributed equally to this paper}
	\affiliation{Department of Physics and Beijing Key Laboratory of
		Opto-electronic Functional Materials $\&$ Micro-nano Devices, Renmin
		University of China, Beijing 100872, P. R. China}
	\author{Changhua
		Bao}\thanks{These authors contributed equally to this paper}
	\affiliation{State Key Laboratory of Low Dimensional Quantum Physics
		and Department of Physics, Tsinghua University, Beijing 100084, P. R. China}
	\author{Yi-Yan Wang}\thanks{These authors contributed equally
		to this paper} \affiliation{Department of Physics and Beijing Key
		Laboratory of Opto-electronic Functional Materials $\&$ Micro-nano
		Devices, Renmin University of China, Beijing 100872, P. R. China}
	\author{Peng-Jie Guo}\thanks{These authors contributed equally to this paper}
	\affiliation{Department of Physics and Beijing Key Laboratory of
		Opto-electronic Functional Materials $\&$ Micro-nano Devices, Renmin
		University of China, Beijing 100872, P. R. China}
	\author{Qiao-He Yu}
	\affiliation{Department of Physics and Beijing Key Laboratory of
		Opto-electronic Functional Materials $\&$ Micro-nano Devices, Renmin
		University of China, Beijing 100872, P. R. China}
	\author{Lin-Lin Sun}
	\affiliation{Department of Physics and Beijing Key Laboratory of
		Opto-electronic Functional Materials $\&$ Micro-nano Devices, Renmin
		University of China, Beijing 100872, P. R. China}
	\author{Yuan Su}
	\affiliation{Department of Physics and Beijing Key Laboratory of
		Opto-electronic Functional Materials $\&$ Micro-nano Devices, Renmin
		University of China, Beijing 100872, P. R. China}
	\author{Kai Liu}
	\affiliation{Department of Physics and Beijing Key Laboratory of
		Opto-electronic Functional Materials $\&$ Micro-nano Devices, Renmin
		University of China, Beijing 100872, P. R. China}
	\author{Zhong-Yi Lu}
	\affiliation{Department of Physics and Beijing Key Laboratory of
		Opto-electronic Functional Materials $\&$ Micro-nano Devices, Renmin
		University of China, Beijing 100872, P. R. China}
	\author{Shuyun
		Zhou} \affiliation{State Key
		Laboratory of Low Dimensional Quantum Physics and Department of
		Physics, Tsinghua University, Beijing 100084, P. R. China}
	\affiliation{Collaborative Innovation Center of Quantum Matter,
		Beijing, P. R. China}
	\author{Tian-Long Xia}\email{tlxia@ruc.edu.cn}
	\affiliation{Department of Physics and Beijing Key Laboratory of
		Opto-electronic Functional Materials $\&$ Micro-nano Devices, Renmin
		University of China, Beijing 100872, P. R. China}
\date{\today}
\begin{abstract}
Quantum limit is quite easy to achieve once the band crossing exists exactly at the Fermi level ($E_F$) in topological semimetals. In multilayered Dirac fermion system, the density of Dirac fermions on the zeroth Landau levels (LLs) increases in proportion to the magnetic field, resulting in intriguing angle- and field-dependent interlayer tunneling conductivity near the quantum limit. BaGa$_2$ is an example of multilayered Dirac semimetal with anisotropic Dirac cone close to $E_F$, providing a good platform to study its interlayer transport properties. In this paper, we report the negative interlayer magnetoresistance (NIMR, I//c and B//c) induced by the tunneling of Dirac fermions on the zeroth LLs of neighbouring Ga layers in BaGa$_2$. When the field deviates from the c-axis, the interlayer resistivity $\rho_{zz}(\theta)$ increases and finally results in a peak with the field perpendicular to the c-axis. These unusual interlayer transport properties (NIMR and resistivity peak with B$\perp$c) are observed together for the first time in Dirac semimetal under ambient pressure and are well explained by the model of tunneling between Dirac fermions in the quantum limit.

\end{abstract}
\pacs{75.47.-m, 71.30.+h, 72.15.Eb}
\maketitle
\setlength{\parindent}{1em}

Topological semimetals have attracted tremendous attention due to their novel properties such as high carrier mobility, large transverse magnetoresistance (MR), non-trivial Berry phase and chiral anomaly\cite{novoselov2005two,zhang2005experimental,wang2012dirac,liu2014discovery,xiong2015evidence,xiong2016anomalous,liu2014stable,borisenko2014experimental,liang2015ultrahigh,he2014quantum,li2015giant,li2016negative,weng2015weyl,xu2015discovery,huang2015weyl,lv2015experimental,lv2015observation,xu2015discovery,xu2015experimental,xu2016observation,liu2016evolution,huang2015observation,zhang2016signatures,
arnold2016negative,sun2015prediction,wang2016mote,tamai2016fermi,deng2016experimental,jiang2017signature,huang2016spectroscopic,zhang2016signatures,yang2015weyl,hu2015pi,borrmann2015extremely}. Quantum oscillations, from which the Berry phase, quantum mobility and effective mass can be extracted, is a routine method in the transport measurement to study topological materials. Obvious Shubnikov-de Haas (SdH) oscillation has been observed in graphene and 3D topological semimetals, e.g.  Cd$_3$As$_2$\cite{borisenko2014experimental,liang2015ultrahigh,he2014quantum,li2015giant,li2016negative,liu2014stable}, Na$_3$Bi\cite{wang2012dirac,liu2014discovery,xiong2015evidence,xiong2016anomalous}, TaAs family\cite{weng2015weyl,xu2015discovery,huang2015weyl,lv2015experimental,lv2015observation,xu2015discovery,xu2015experimental,xu2016observation,liu2016evolution,huang2015observation,zhang2016signatures,
arnold2016negative,zhang2016signatures,yang2015weyl,hu2015pi,borrmann2015extremely} and MoTe$_2$\cite{sun2015prediction,wang2016mote,tamai2016fermi,deng2016experimental,jiang2017signature,huang2016spectroscopic}.  A particular interesting case is when the quantum limit is achieved, only the zeroth LL is occupied and no more quantum oscillations should be observed. With the increase of the field, the increased zeroth LL's degeneracy leads to the monotonic increase of density of state, inducing exotic phenomena in multilayered Dirac/Weyl materials e.g. negative longitudinal interlayer magnetoresistance and a peak in the angle-dependent interlayer resistivity as reported in $\alpha$-(BEDT-TTF)$_2$I$_3$ (under pressure) and YbMnBi$_2$(NIMR not observed)\cite{osada2008negative,tajima2009effect,liu2016zeeman,liu2017unusual}. $\alpha$-(BEDT-TTF)$_2$I$_3$ is a typical multilayer system which consists of conductive layers of BEDT-TTF molecules and insulating layers of I$_3$$^-$. In $\alpha$-(BEDT-TTF)$_2$I$_3$, the Dirac point is located exactly at $E_F$  under pressure. As for YbMnBi$_2$, the Weyl points are believed to originate from the 2D Bi square-net planes and they are also located at the $E_F$\cite{borisenko2015time,liu2017unusual}.\\

\begin{figure}
\centering
  \includegraphics[width=0.48\textwidth]{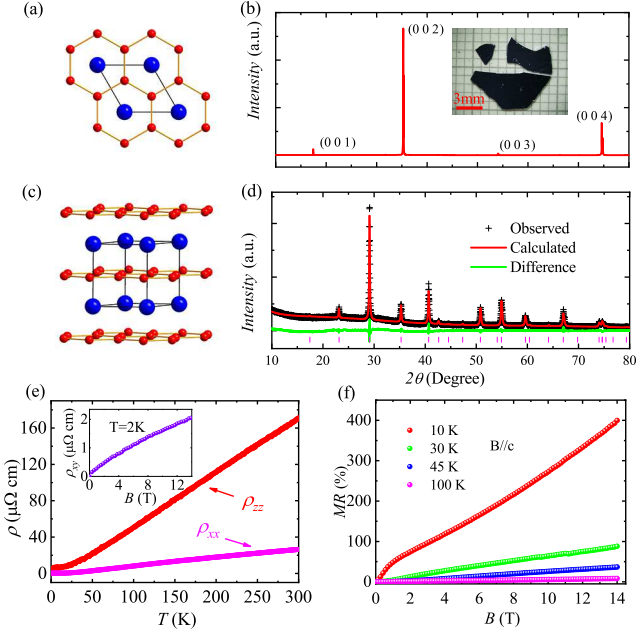}\\
  \caption{(color online) (a)(c) Crystal structure of BaGa$_2$. The red balls represent the Ga, and the blue balls represent Ba. (b) Single crystal X-ray diffraction pattern. Insert: the picture of BaGa$_2$ crystal. (d) Powder X-ray diffraction pattern with refinement by TOPAS, $a = 4.43{\AA}$, $c = 5.08{\AA}$. (e) In-plane $\rho_{xx}$(T) and interlayer $\rho_{zz}$(T) versus T of BaGa$_2$. Insert: Hall resistivity $\rho_{xy}$ versus magnetic field at 2 K. (f) In plane MR at different temperatures. }\label{2}
\end{figure}

\begin{figure*}
\centering
  \includegraphics[width=0.8\textwidth]{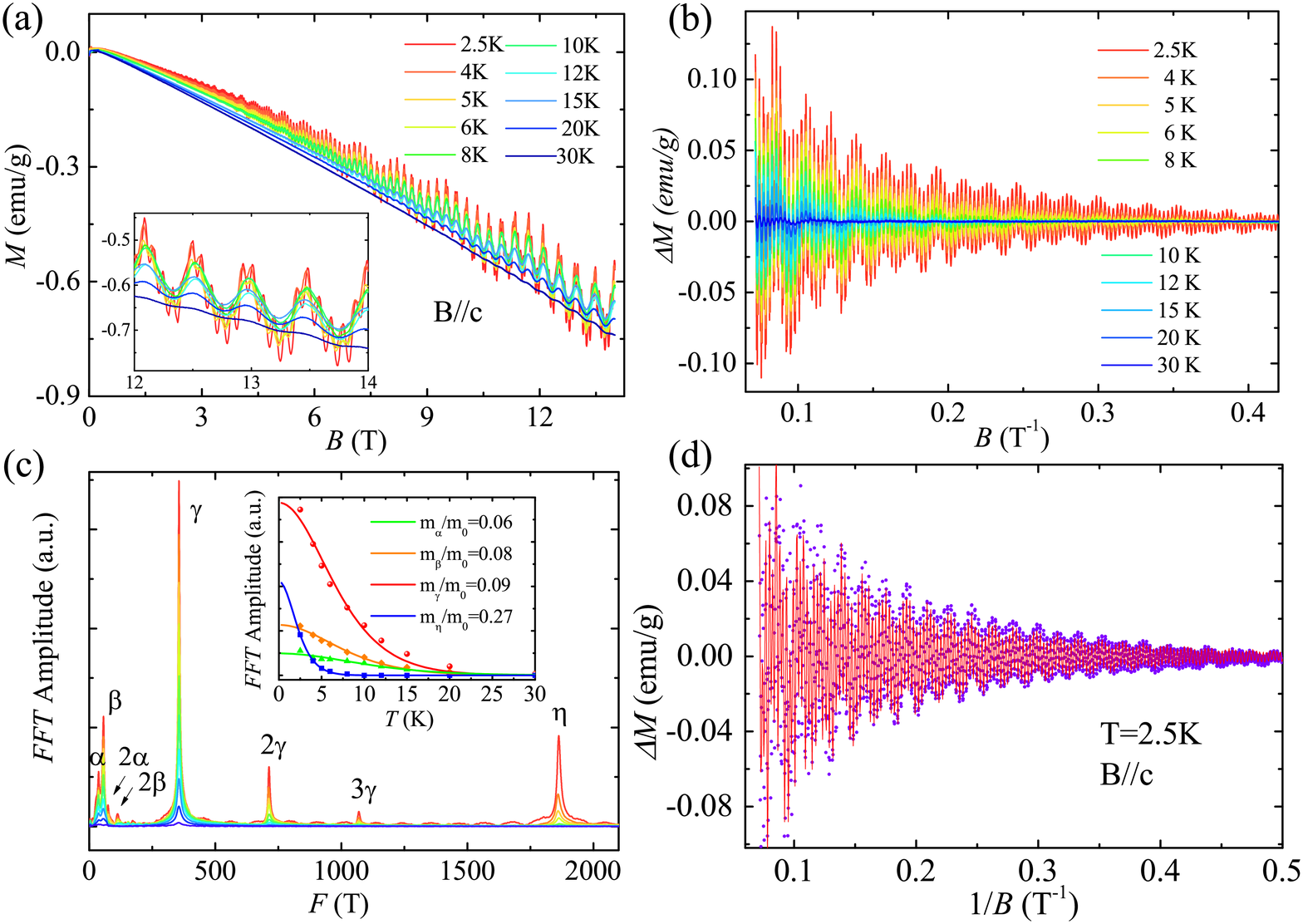}\\
  \caption{(color online) (a)The dHvA oscillation for \emph{B}//\emph{c}. (b) The oscillatory component \emph{$\Delta$M} as a function of 1/\emph{B} after subtracting a smoothing background. (c) FFT spectra of the \emph{$\Delta$M}. Inset: the effective masses fitted by the  thermal factor of LK formula. (d) The LK fit (red line) of \emph{$\Delta$M} at 2.5 K.}\label{2}
\end{figure*}

\begin{figure}
\centering
  \includegraphics[width=0.48\textwidth]{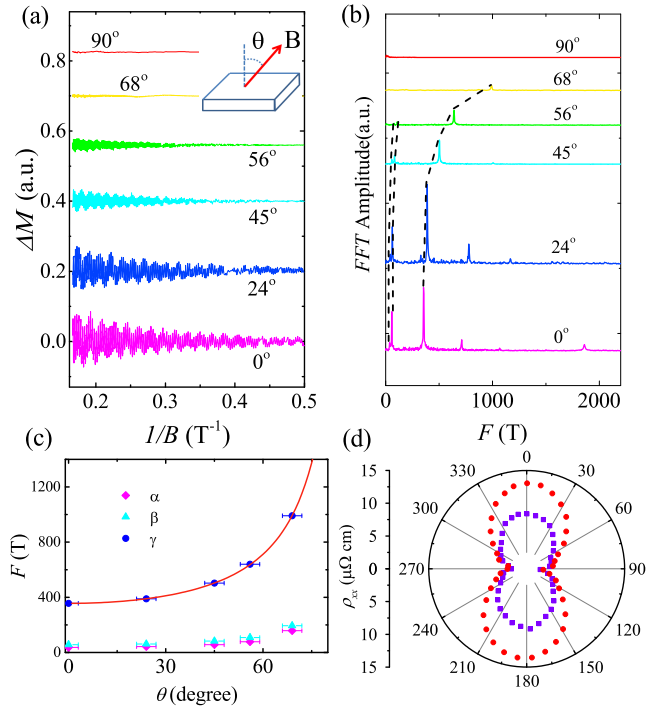}\\
  \caption{(color online) (a) Angle-dependent dHvA oscillations at 1.8K. (b) Corresponding FFT spectra for $\theta$ varying from 0$^\circ$ to 90$^\circ$. (c) FFT frequencies as a function of angle. Solid red line is the fitting line with equation $F = F_0/cos\theta$, F$_0$ equals to 356T for the $\gamma$ pocket. Error bars derive from our measurement deviation. (d) Polar plot of MR at 9 T (violet dots) and 14 T (red dots) with B always vertical to I at 2.5 K.}\label{2}
\end{figure}

\begin{figure}
\centering
  \includegraphics[width=0.48\textwidth]{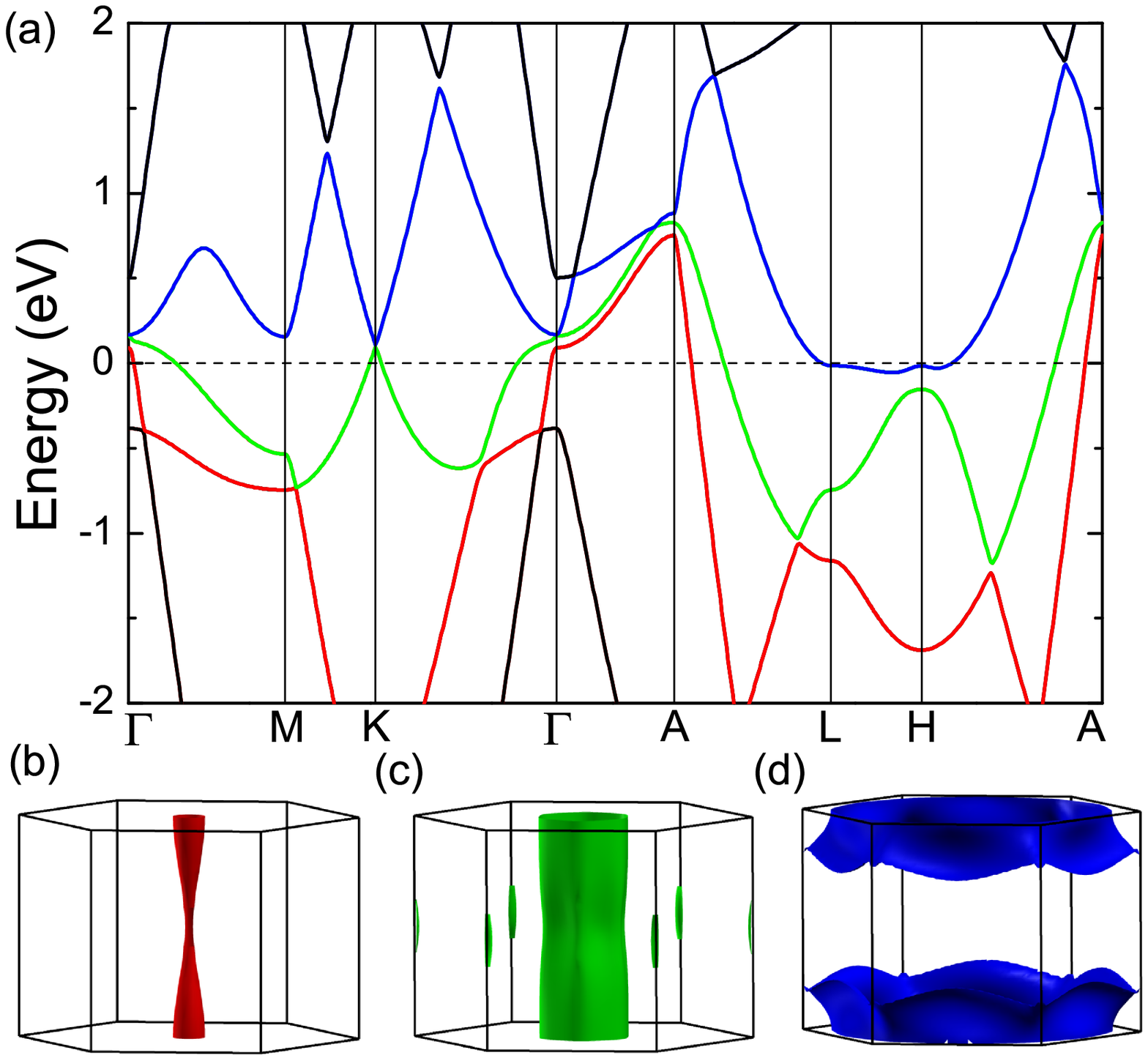}\\
  \caption{(color online) (a) Band structure of BaGa$_2$ along high-symmetry paths of the Brillouin zone calculated with SOC. (b)-(c) Two hole-type Fermi surfaces and (d) one electron-type Fermi surface of BaGa$_2$.}\label{1}
\end{figure}

BaGa$_2$ is recently predicted to be a candidate of Dirac semimetals, in which the quasi-2D Dirac cone comes from Ga $p_z$ orbit with the Dirac point located at \emph{K} \cite{gibson2015three}. Since the Dirac point is located at almost exactly the $E_F$, the quantum limit can be easily reached when a relatively small magnetic field is applied. Thus, the zeroth LL's degeneracy increases with the field and the Dirac fermions in the zeroth LL's participate in the magneto-transport directly, which provides a good platform to investigate the unusual interlayer transport properties induced by the tunneling of Dirac fermions. In this work, the quasi-2D Dirac cone at \emph{K} with the Dirac point located at E$_F$ is confirmed by the first-principles calculations, angle-dependent dHvA oscillations and ARPES measurements. The above-mentioned NIMR and peak in angle-dependent interlayer resistivity are both observed clearly in BaGa$_2$ for the first time and well explained by the model of zeroth LL's Dirac fermion tunneling\cite{osada2008negative}.

\section{Materials characteristic and in-plane transport properties}

BaGa$_2$ single crystals are grown by self-flux method (see the method). The crystal structure consists of Ba square-net planes and the Ga honeycomb-net planes which shows 2D characteristic. The (\emph{00l}) lattice plane and lattice parameters are determined by the X-ray diffraction (XRD) measurements on BaGa$_2$ single crystal and powder  as shown in Figs. 1(b) and 1(d). Both the in-plane resistivity ($\rho_{xx}$) and the interlayer resistivity ($\rho_{zz}$) show metallic behavior [Fig.1(e)]. The ratio of $\rho_{zz}$/$\rho_{xx}$ is about 26 at 2.5 K, indicating the anisotropic electronic structure. The measurements of Hall resistivity $\rho_{xy}$ reveal that the dominant carrier is hole with the carrier concentration and mobility estimated to be $n_h=4.28\times 10^{21} cm^{-3}$ and $\mu_h=3277 cm^2/Vs$  at 2 K. The MR of BaGa$_2$ reaches 400$\%$ and shows no signs of SdH oscillation [Fig.1(f)].\\

\section{dHvA quantum oscillations}

Figure. 2(a) presents the intense dHvA oscillations at various temperatures with the magnetic field along [\textit{00l}] direction. Four fundamental frequencies $F_\alpha$=36.9 T, $F_\beta$=56.6 T, $F_\gamma$=356 T, $F_\eta$=1862 T are confirmed to exist in the oscillations after fast Fourier transform (FFT) analysis of the oscillatory component \emph{$\Delta$M} as shown in Fig. 2(c). According to the Onsager relation, the four frequencies correspond to four cross sections of the Fermi surfaces (FSs).

\begin{figure*}
\centering
    \includegraphics[width=0.9\textwidth]{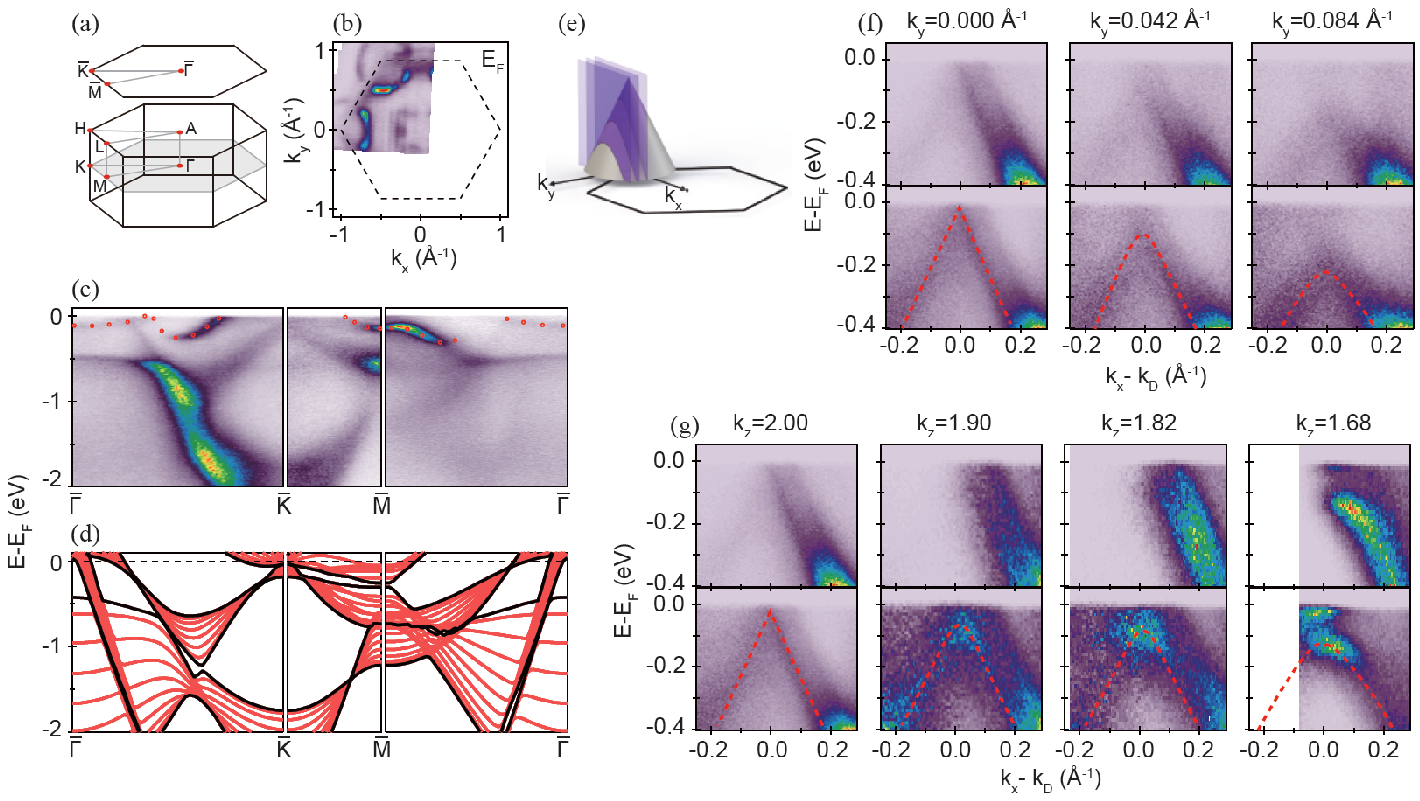}\\
  \caption{(color online) Fermi surface, electronic dispersions and three-dimensional Dirac cone in BaGa$_2$. (a) Schematic drawing of the 3D Brillouin zone with high symmetry points marked by red points and its projected surface Brillouin zone to (001) surface. (b) Measured Fermi surface with corresponding Brillouin zone. (c) Measured electronic dispersion along $\overline{\Gamma}-\overline{K}-\overline{M}-\overline{\Gamma}$ direction. (d) Calculated bulk electronic dispersion with different $k_z$ along $\overline{\Gamma}-\overline{K}-\overline{M}-\overline{\Gamma}$ direction. (e) Schematic drawing shows the slices (purple plane) which cut through the Dirac cones at different $k_y$. (f) Electronic dispersions cut through the Dirac cones at corresponding $k_y$ and corresponding electronic dispersions in the upper panel after normalization by integrating energy distribution curve (EDC). (g) $k_z$ dependence of electronic dispersions cut through the Dirac point and corresponding electronic dispersions in the upper panel after normalization by integrating EDC. The unit of $k_z$ is $2\pi/c$.}\label{5}
\end{figure*}

The dHvA oscillations can be described by the Lifshitz-Kosevich (LK) formula\cite{shoenberg2009magnetic},
\begin{equation}\label{equ1}
\Delta M \propto\ -B^\gamma  R{_T}R{_D}sin\left[2\pi\times\left(\frac{F}{B}-\frac{1}{2}+\beta+\delta\right)\right]
\end{equation}
where $R{_T}=(\lambda T\mu/B) / sinh(\lambda T\mu/B)$, $R{_D}=exp(-\lambda T{_D}\mu/B)$ and  $\lambda=(2\pi^{2} k{_B} m{_0})/(\hbar e)$. $\mu$ is the ratio of effective mass $m^*$ to free electron mass $m_0$, and $T{_D}$ is the Dingle temperature.  $\gamma=0$, $\delta=0$ for a 2D system, and $\gamma=1/2$, $\delta=\pm1/8$ for a 3D system. $\beta=\phi_B/2\pi$ and $\phi_B$ is the Berry phase. The inset of Fig. 2(c) displays the temperature-dependent FFT amplitudes  and fittings using the thermal factor $R{_T}$ in LK formula. The obtained effective cyclotron masses are quite small, comparable as those in topological semimetals, due to the almost linear dispersion though they all originate from trivial bands lately determined by the magnetotransport analysis and first-principles calculations. In order to extract the Berry phase of BaGa$_2$ conveniently, we applied four-band LK formula fitting directly. The fitting result is displayed in Fig. 2(d), the violet dots are experimental results and the red line is the fitting curve. All of the Berry phase of these four pockets greatly deviate from the non-trivial value $\pi$ implying that these bands may be trivial. Detailed data is exhibited in Table I. The angle-dependent dHvA oscillation measurements are applied to further reveal the Fermi surface of BaGa$_2$ as exhibited in Fig.3(a). Figure. 3(c) shows the FFT spectra of the corresponding dHvA oscillations with the B rotating from B//c to B//ab. The values of fundamental frequencies increase with the angle $\theta$ and vanish at $\theta=90^\circ$, indicating the quasi-2D characteristics of the Fermi surfaces. Besides, twofold anisotropy polar plot MR [Fig. 3(d)] with B always vertical to I at 2.5 K also implies the quasi-2D characteristics of the Fermi surfaces.\\

\begin{table*}
  \centering
\caption{Parameters of BaGa$_2$ calculated from dHvA oscillations. \emph{F} is frequency of the dHvA oscillations; $T_D$ is Dingle temperature; $m^*$/$m_0$ is the ratio of the carrier effective mass to bare electron mass; $\tau$$_q$ is quantum relaxation time; $\mu$$_q$ is quantum mobility; $\phi$$_B$ represents Berry phase; $\upsilon$$_F$, $k_F$; and $E_F$ represent Fermi velocity, Fermi vector and Fermi energy, respectively.}
\tabcolsep 0.08in
\renewcommand\arraystretch{1.5}
\begin{tabular}{ccccccccccc}
\hline
peak & \emph{F} (T) & $T_D$ (K) & $m^*$/$m_0$ & $\tau$$_q$(ps) & $\mu$$_q$(cm$^2$/Vs) & $\phi$$_B$  & $\upsilon$$_F$ ($10^6$ m/s)  & $k_F$($\AA^{-1}$) & $E_F$(eV)\\
\hline
$\alpha$ & 36.9 & 27.3 & 0.06 & 0.045 & 1334 & 1.39$\pi$  & 0.66 & 0.033 & 0.144 \\
$\beta$  & 56.6 & 8.0 & 0.08 & 0.152 & 3424 & 1.73$\pi$  & 0.62 & 0.041 & 0.168\\
$\gamma$ & 356 & 5.9 & 0.09 & 0.21 & 4107 & 0.75$\pi$   & 1.37 & 0.104 & 0.937\\
$\eta$ & 1862 & 7.2 & 0.27 & 0.168 & 1179 & 1.37$\pi$  & 1.02 & 0.238 & 1.597\\
\hline
\end{tabular}
\end{table*}

\section{First-principles calculations}

The band structure and Fermi surfaces of BaGa$_2$ calculated with spin-orbit coupling (SOC) effect included are shown in Fig. 4. There exist three bands crossing the Fermi level with the corresponding FSs demonstrated in Figs. 4(b) - 4(d). The hole-type FSs (Figs. 4(b) and 4(c)) are open-orbit FSs along the $k_z$ direction and exhibit strong two-dimensional characteristic, which is consistent with the angle-dependent dHvA oscillations (Fig.3). Notably, the hole-type FS in Fig. 4(b) has two evident cross sections corresponding to the frequencies $F_\alpha$ and $F_\beta$. The frequencies $F_\gamma$ and $F_\eta$ originate from the FSs in Figs. 4(c) and 4(d), respectively. Thus, the trivial Berry phase obtained from the fitting can be understood, because these four frequencies are all from the trivial bands. The Dirac point locates at the high-symmetry point \emph{K}. Although the Dirac point is not protected by the crystal symmetry and opens a small gap of $\sim$ 10 meV by considering the SOC effect, the negligible gap has little effect on the transport properties. Next, ARPES is employed to confirm the Dirac cone at K points.\\

\section{ARPES measurement}
The electronic structure of BaGa$_2$ is measured by ARPES and the quasi-2D Dirac cone at the K point is revealed. Figure 5(a) shows a schematic drawing of the 3D Brillouin zone (BZ) and its projected surface BZ onto the (\emph{00l}) surface measured. The Fermi surface is shown in Fig. 5(b). The most obvious feature is a nearly circular pocket around the $\overline{K}$ point. Besides this pocket, there are several much weaker features around the $\overline{\Gamma}$ and $\overline{K}$ points where several bands cross the Fermi energy. This can be seen more clearly in the ARPES spectra along the high-symmetry directions as shown in Fig. 5(c). The experimental spectra show an overall agreement with the calculated dispersions at different $k_z$ values shown in Fig. 5(d), except for a few extra band dispersions near the Fermi energy (indicated by red circles in Fig. 5(c)). These extra dispersions may originate from surface states which are not included in the calculation. Besides, the experimental dispersions which are away from Fermi energy show a mixture of calculated dispersions ranging from $k_z=0$ and $k_z=0.5c^*$ (the dispersions at the $k_z$ boundaries are colored black in Fig. 5(d)), suggesting a very strong $k_z$-broadening \cite{Kz}.

We zoom in the dispersion near $E_F$ to further reveal the Dirac cone at the \emph{K} point. The dispersion at $k_y$=0 is shown in Fig. 5(f) . Due to the matrix element effect \cite{ARPESRev}, the left branch is much weaker than the right branch. To enhance the visualization of the dispersion on both sides, we show in Fig. 5(g) the normalized spectra by integrating each energy distribution curve (EDC). In addition to the Dirac cone at the \emph{K} point, there is an extra dispersion outside the Dirac cone. This may originate from the surface state. The evolution of Dirac cone along the $k_y$ direction is shown in Fig. 5(f). Moving away from the \emph{K} point, the dispersion shifts down in energy and becomes more parabolic, in agreement with  a Dirac cone at the \emph{K} point as shown in Fig. 5(a). The $k_z$-dependence of the Dirac cone is  shown in Fig. 5(g). Similar behavior is also observed, with the dispersion shifting down in energy when moving away from \emph{K} point ($k_z=2c^*$), suggesting that this Dirac cone is quasi-2D and centered at \emph{K} point.\\

\begin{figure*}
\centering
  \includegraphics[width=0.9\textwidth]{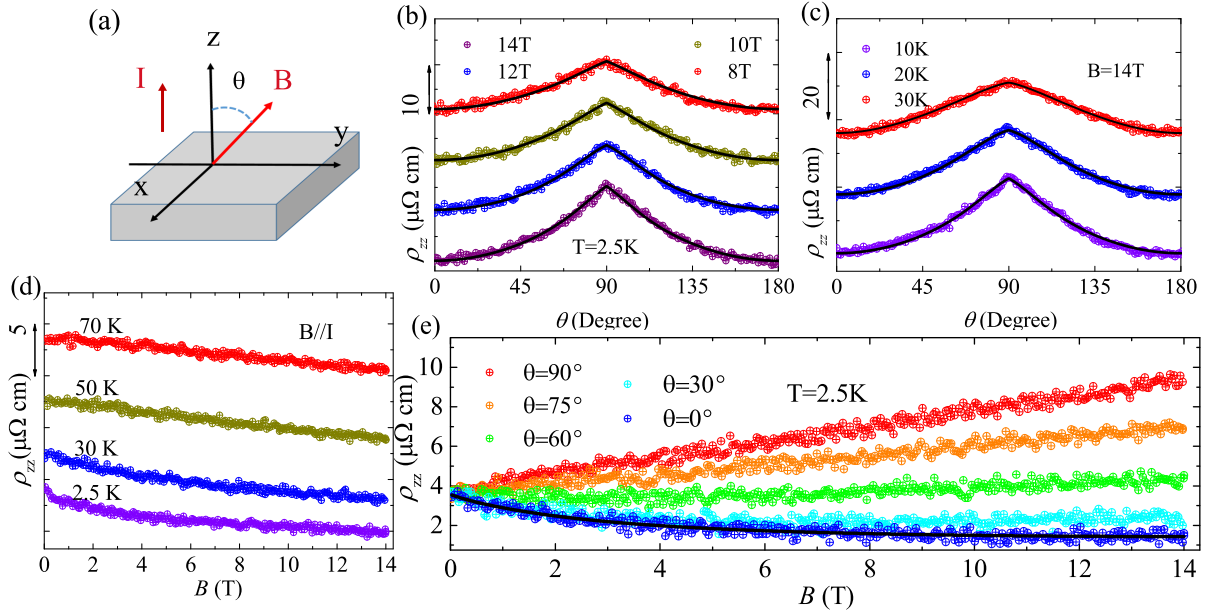}\\
  \caption{(color online) (a) The interlayer transport experiment setup, \emph{I} flow along \emph{c}, $\theta$ is the angle between \emph{B} and \emph{I}. (b) Angle dependence of interlayer resistivity $\rho_{zz}$ at 2.5 K and different magnetic fields. (c) Angle dependence of $\rho_{zz}$ at 14 T and various temperatures. The black curves covered on the data in (b) is the fits using Eq. (2). (d)(e) Field dependence of $\rho_{zz}$ at various temperatures and sample positions.}\label{1}
\end{figure*}

\section{Interlayer transport properties}
BaGa$_2$ provides an ideal platform to study the unusual interlayer transport properties caused by the tunneling of Dirac fermions on the zeroth LLs. Figures. 6(b) and 6(c) present the angle-dependent interlayer resistivity $\rho_{zz}(B,\theta)$ at different fields and temperatures ($\theta$ is the angle between \emph{\textbf{B}} and \emph{\textbf{I}} as defined in Fig. 6(a)). The curves of $\rho_{zz}(B,\theta)$ show the interlayer resistivity peak when $\theta=90^\circ$. Besides, when $\theta=0^\circ$, the interlayer resistivity of BaGa$_2$ decreases with the increasing of field, resulting in the negative interlayer MR as shown in Fig. 6(d). The NIMR does not vanish with the temperature up to 70 K. Both of these anomalous interlayer transport properties can be understood based on the tunneling of Dirac fermions on the zeroth LLs. According to previous works\cite{osada2008negative,liu2017unusual}, the total interlayer resistivity $\rho_{zz}(B,\theta)$ can be described as
\begin{equation}\label{equ1}
\rho_{zz}(B,\theta) \approx \frac{1}{\sigma_{zz}(B,\theta)}=\frac{1}{\sigma_t^{LL0}(B,\theta)+\sigma_c(B,\theta)+B_0}
\end{equation}
Where $\sigma_c(B,\theta)$ is the conductivity from trivial bands. As shown in the Supplementary table I, all of the trivial bands do not reach the quantum limit, so the conductivity $\sigma_c(B,\theta)$ is taken as $\sigma_0/(1+k\cdot B_{xy}^2)$ for low field approximation ($\sigma_0$ is the Drude conductivity, \emph{k} is a constant, $B_{xy}=|\emph{B}|sin\theta$). $B_0$ is a fitting parameter\cite{osada2008negative}. $\sigma_t^{LL0}(B,\theta)$ represents the tunneling conductivity of Dirac fermions and can be taken as $A\left|Bcos\theta\right|exp[-\frac{1}{2}\frac{ec^2(Bsin\theta)^2}{\hbar\left|Bcos\theta\right|}]$ ($\theta$ is the angle between \emph{B} and \emph{I}, \emph{c} is the distance of the adjacent two Ga layers). When the field is parallel to \emph{I} along \emph{c}-axis ($\theta=0^\circ$), the tunneling conductivity of Dirac fermions can be simplified as $\sigma_t^{LL0}(B,0^\circ)$=A$\left|B\right|$, which increases in proportion to the magnetic field. Thus, the interlayer resistivity $\rho_{zz}(B, 0^{\circ})$ decreases with the  magnetic field, resulting in the NIMR. The field-dependent interlayer resistivity can be well fitted by Eq. (2) as shown in Fig. 6(e). The interlayer resistivity in magnetic field can be considered as a competition between the $B$-dependent tunneling conductivity and the $B^{-2}$-dependent trivial band conductivity with $\theta$ varying from 0$^\circ$ to 90$^\circ$. When $\theta$ is close to 0$^\circ$, the tunneling conductivity is dominant, accompanying with the NIMR. When $\theta$ is close to 90$^\circ$, the trivial band dominates the interlayer conductivity, leading to the positive magnetoresistance. Thus, the NIMR disappeared gradually when $\theta$ increases from 0$^\circ$ to 90$^\circ$ as shown in Fig. 6(e). When $\theta=90^\circ$, the tunneling conductivity $\sigma_t^{LL0}(B,90^\circ)=0$ and the interlayer resistivity peak appears, which is well fitted by Eq. (2) (Figs. 6(b) and 6(c)). These anomalous interlayer transport properties induced by the tunneling of  Dirac fermions on the zeroth LLs have also been observed in (BEDT-FFT)$_2$I$_3$, where  the Dirac cone only exist under pressure. YbMnBi$_2$ is another quasi-2D topological material that exhibits the same anomalous interlayer transport properties\cite{liu2017unusual}, in which the NIMR was not observed possibly due to the large positive MR induced by other bands overwhelming  the signature of NIMR. To our knowledge, BaGa$_2$ is the only material that exhibits both NIMR and interlayer resistivity peak at ambient pressure.\\

\section{Discussions}
To further understand the observed NIMR in BaGa$_2$, we discuss on the possibly different origins of the negative magnetoresistance (NMR) below. BaGa$_2$ is a non-magnetic material, spin flip may  not be a suitable reason for the NIMR in BaGa$_2$\cite{ritchie2003magnetic}. Secondly, the NMR has also been observed in topologically trivial materials at high field such as graphite, where the NMR is induced by the  ellipsoidal Fermi surfaces approaching the quantum limit\cite{fauque2013two}. As displayed in the table I of the Supplementary, all of the trivial bands in BaGa$_2$ are not under the quantum limit with the field of 14 T. In addition, the NIMR in BaGa$_2$ appears once a quite small field is applied. Thus, the scenario of trivial bands under quantum limit does not apply in BaGa$_2$. Thirdly, the NMR induced by chiral anomaly has been widely observed in 3D Dirac and Weyl semimetals\cite{xiong2015evidence,liang2015ultrahigh,li2016negative,huang2015observation,zhang2016signatures}. The NMR induced by chiral anomaly usually has the form of  $\Delta\sigma_{xx}(B)$$\propto$$B^2$ or $B$ depending on the Fermi energy\cite{dai2016detecting}. However, the in-plane NMR is not observed when the field is parallel to the current in BaGa$_2$ (see Supplementary Fig. 1). So the scenario of chiral anomaly induced NMR does not apply either. Fourthly, current jetting can often cause the NMR\cite{pippard1989magnetoresistance,hu2005current}. Thus, we have shaped the sample and repeatedly improved the electrode to avoid the geometry and size effect (see Supplementary Fig. 2). Finally, the NIMR can arise in high-purity layered metal PdCoO$_2$ with the residual resistivity $\rho_{xx}$ ranging from about 10 to 40 n$\Omega \cdot cm$ in most single crystals\cite{kikugawa2016interplanar}. As for BaGa$_2$, the residual resistivity $\rho_{xx}$ and $\rho_{zz}$ are about 440 n$\Omega \cdot cm$ and 11.44 u$\Omega \cdot cm$ which are larger than that of PdCoO$_2$ at least one order of magnitude. So the condition $4t_c >\hbar\omega_c$ ($\varepsilon_n=(n+1/2)\hbar \omega_c +2t_c cos(k_zc)$ describes the Landau levels for quasi-2D Fermi surface, where t$_c$ is interlayer transfer integral) under the magnetic field in the quasi-2D metal material PdCoO$_2$   may not be fulfilled here. Besides, the interlayer MR in PdCoO$_2$ increases quickly at low field and decreases at high field when field deviates from c-axis which is quite different from that of BaGa$_2$. From the above, we conclude that the interlayer NMR and the interlayer resistivity peak result from the tunneling of zeroth LLs Dirac fermions.

\section{Summary}

In summary, we have grown the single crystal of BaGa$_2$ and studied its electronic structure and transport properties systematically. Angle-dependent dHvA quantum oscillations are consistent with the first-principles calculations which confirmed the quasi-2D characteristic of the Fermi surface in BaGa$_2$. The first-principles calculations together with the ARPES measurement confirmed the quasi-2D Dirac cone located at K point. The dHvA oscillations originate from the trivial bands while absent for the Dirac band near the Fermi level. BaGa$_2$ exhibits negative interlayer magnetoresistance, and the angle-dependent interlayer resistivity $\rho_{zz}(B, \theta)$ shows a peak at $\theta$=90$^\circ$. Both of these unusual interlayer transport properties originate from the zeroth LL's Dirac fermions tunneling, and can be well fitted by the tunneling model. The field and angle-dependent characteristics in BaGa$_2$ provide us a new platform to further study the potential applications in the future quantum devices.

\section{Method}

The single crystal BaGa$_2$ was grown by self-flux method. The starting ingredient ratio with Ba:Ga=41.7:58.3 were put into the alumina crucible and sealed into a quartz ampoule. Then the ampoule was heated to 950 $^\circ$C and kept for 50 h to make sure the materials have melted and mixed thoroughly. The process of crystal growth was carried up in cooling at 3 $^\circ$C/h to 700 $^\circ$C. The BaGa$_2$ single crystal was obtained after centrifuging the excess flux at that temperature. The atomic composition of BaGa$_2$ was checked by energy dispersive x-ray spectroscopy (EDS, Oxford X-Max 50). The crystal structure was determined by X-ray diffraction (XRD) in Bruker D8 Advance x-ray diffractometer. TOPAS-4.2 was employed for refinement. The electric transport and magnetic properties were collected in Quantum Design PPMS-14 T and MPMS-7 T SQUID VSM system. ARPES measurements were taken at BL13U of Hefei National Synchrotron Radiation Laboratory and our home laboratory. The crystals were cleaved in situ and measured at a temperature of T $\approx$ 15 K in vacuum with a base pressure better than  $1\times 10^{-10}$ torr.

The electronic structures of BaGa$_2$ were studied by using first-principles calculations. The projector augmented wave (PAW) method \cite{PhysRevB.50.17953,PhysRevB.59.1758} as implemented in the VASP package\cite{PhysRevB.47.558,kresse1996,PhysRevB.54.11169} was used to describe the core electrons. For the exchange-correlation potential, the generalized gradient approximation (GGA) of Perdew-Burke-Ernzerhof formula\cite{PhysRevLett.77.3865} was adopted. The kinetic energy cutoff of the plane-wave basis was set to be 400 eV. A 12$\times$12$\times$10 \emph{k}-point mesh was utilized for the Brillouin zone (BZ) sampling and the Fermi surface was broadened by the Gaussian smearing method with a width of 0.05 eV. Both cell parameters and internal atomic positions were fully relaxed until the forces on all atoms were smaller than 0.01 eV/A. Once the equilibrium structures were obtained, the electronic structures were calculated with the spin-orbital coupling (SOC) effect. The Fermi surfaces were studied by using the maximally localized Wannier functions (MLWF) method\cite{PhysRevB.56.12847,PhysRevB.65.035109}. 

\section{Acknowledgments}
The authors would like to thank T. Qian, G. F. Chen and H. C. Lei for helpful discussions. This work is supported by the National Natural Science Foundation of China (No.11574391, No.11725418, No. 11774422, No. 11774424, No.11874422), the National Key R\&D Program of China (Grant No. 2017YFA0302903) and the Fundamental Research Funds for the Central Universities, and the Research Funds of Renmin University of China (RUC) (No.18XNLG14, No.19XNLG13, No.19XNLG18). Computational resources were provided by the Physical Laboratory of High Performance Computing at Renmin University of China. The Fermi surfaces were prepared with the XCRYSDEN program\cite{kokalj2003computer}.
\bibliography{bibtex}

\end{document}


Supplemental Online Material
	\bibliographystyle{apsrev}
	\title{Interlayer quantum transport in Dirac semimetal BaGa$_2$}
	
	\author{Sheng Xu}\thanks{These authors contributed equally to this paper}
	\affiliation{Department of Physics and Beijing Key Laboratory of
		Opto-electronic Functional Materials $\&$ Micro-nano Devices, Renmin
		University of China, Beijing 100872, P. R. China}
	\author{Changhua
		Bao}\thanks{These authors contributed equally to this paper}
	\affiliation{State Key Laboratory of Low Dimensional Quantum Physics
		and Department of Physics, Tsinghua University, Beijing 100084, P. R. China}
	\author{Yi-Yan Wang}\thanks{These authors contributed equally
		to this paper} \affiliation{Department of Physics and Beijing Key
		Laboratory of Opto-electronic Functional Materials $\&$ Micro-nano
		Devices, Renmin University of China, Beijing 100872, P. R. China}
	\author{Peng-Jie Guo}\thanks{These authors contributed equally to this paper}
	\affiliation{Department of Physics and Beijing Key Laboratory of
		Opto-electronic Functional Materials $\&$ Micro-nano Devices, Renmin
		University of China, Beijing 100872, P. R. China}
	\author{Qiao-He Yu}
	\affiliation{Department of Physics and Beijing Key Laboratory of
		Opto-electronic Functional Materials $\&$ Micro-nano Devices, Renmin
		University of China, Beijing 100872, P. R. China}
	\author{Lin-Lin Sun}
	\affiliation{Department of Physics and Beijing Key Laboratory of
		Opto-electronic Functional Materials $\&$ Micro-nano Devices, Renmin
		University of China, Beijing 100872, P. R. China}
	\author{Yuan Su}
	\affiliation{Department of Physics and Beijing Key Laboratory of
		Opto-electronic Functional Materials $\&$ Micro-nano Devices, Renmin
		University of China, Beijing 100872, P. R. China}
	\author{Kai Liu}
	\affiliation{Department of Physics and Beijing Key Laboratory of
		Opto-electronic Functional Materials $\&$ Micro-nano Devices, Renmin
		University of China, Beijing 100872, P. R. China}
	\author{Zhong-Yi Lu}
	\affiliation{Department of Physics and Beijing Key Laboratory of
		Opto-electronic Functional Materials $\&$ Micro-nano Devices, Renmin
		University of China, Beijing 100872, P. R. China}
	\author{Shuyun
		Zhou} \affiliation{State Key
		Laboratory of Low Dimensional Quantum Physics and Department of
		Physics, Tsinghua University, Beijing 100084, P. R. China}
	\affiliation{Collaborative Innovation Center of Quantum Matter,
		Beijing, P. R. China}
	\author{Tian-Long Xia}\email{tlxia@ruc.edu.cn}
	\affiliation{Department of Physics and Beijing Key Laboratory of
		Opto-electronic Functional Materials $\&$ Micro-nano Devices, Renmin
		University of China, Beijing 100872, P. R. China}
	\date{\today}
	
	\maketitle

\begin{figure*}
 \includegraphics[width=0.7\textwidth]{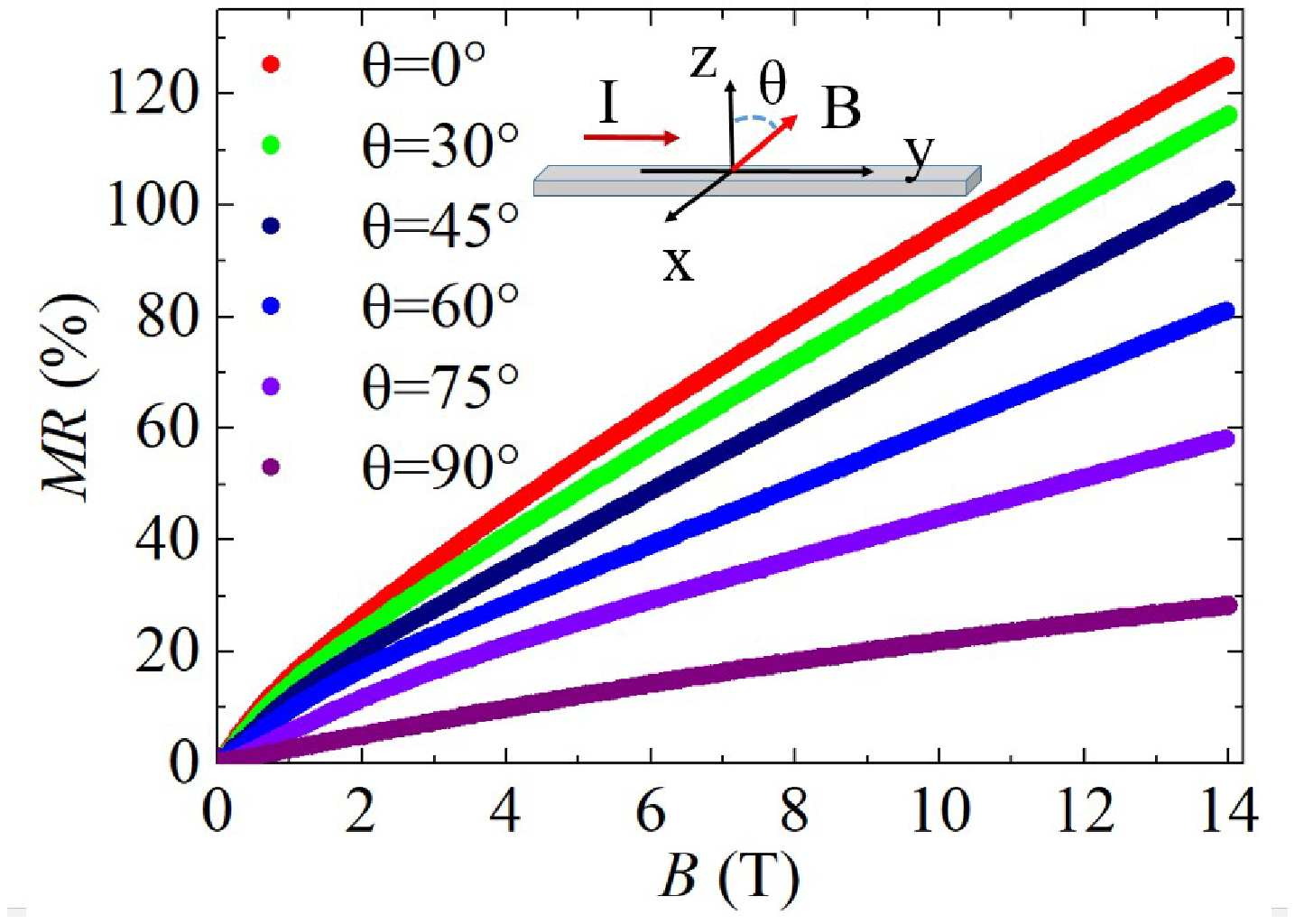}\\
\caption{In-plane MR of BaGa$_2$ under different magnetic field orientations at 2.5 K.}

\end{figure*}
\clearpage

\begin{figure*}
 \includegraphics[width=0.6\textwidth]{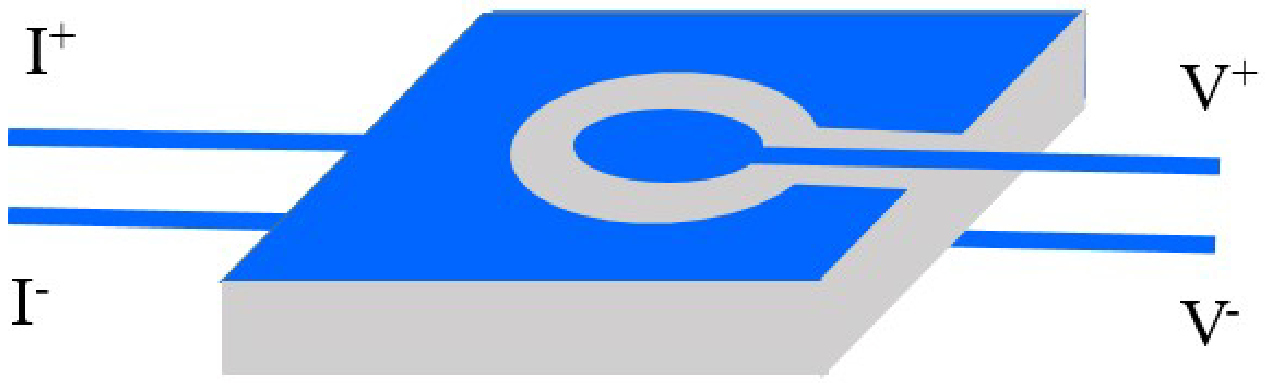}\\
\caption{
 Improved electrode in interlayer transport measurement.
}
\end{figure*}
\clearpage

\begin{table*}
\caption{Landau level index N for the trivial pockets. These parameters are obtained from the dHvA quantum oscillations measurement at 14 T. All of these four pockets are not in the quantum limit, while the negative interlayer MR in BaGa2 appeared at a comparatively small magnetic field which are far away from the quantum limit.}
\tabcolsep 0.5in
\renewcommand\arraystretch{1.5}
\begin{tabular}{ccccc}
\hline
\hline
   & $F_{\alpha}$  & $F_{\alpha}$ & $F_{\alpha}$ & $F_{\eta}$ \\
\hline
LL index N & 3 & 5 & 28 & 134  \\
\hline
\hline
\end{tabular}
\end{table*}
\clearpage
